\documentclass[aps,pre,twocolumn,superscriptaddress,showpacs,showkeys ]{revtex4-1}

\usepackage{graphicx}                   
\usepackage{hyperref}                   
\usepackage{amsmath,amssymb}            
\usepackage{epstopdf}
\usepackage{subcaption}					

\usepackage{color}
\definecolor{orange}{rgb}{1,0.5,0}
\definecolor{brown}{rgb}{0.65, 0.16, 0.16}
\definecolor{phlox}{rgb}{0.87, 0.0, 1.0}

\graphicspath{{figs/}}					

\begin{document}

\title{Gaussian Free Field in the iso-height random islands tuned by percolation model}

\author{J. Cheraghalizadeh}
\affiliation{Department of Physics, University of Mohaghegh Ardabili, P.O. Box 179, Ardabil, Iran}
\email{jafarcheraghalizadeh@gmail.com}

\author{M. N. Najafi*}
\affiliation{Department of Physics, University of Mohaghegh Ardabili, P.O. Box 179, Ardabil, Iran}
\email{morteza.nattagh@gmail.com}

\author{H. Mohammadzadeh}
\affiliation{Department of Physics, University of Mohaghegh Ardabili, P.O. Box 179, Ardabil, Iran}
\email{h.mohammadzadeh@gmail.com}

\begin{abstract}
The Gaussian free field (GFF) is considered in the background of random iso-height islands which is modeled by the site percolation with the occupation probability $p$. To realize GFF, we consider the Poisson equation in the presence of normal distributed white-noise charges, as the stationary state of the Edwards-Wilkinson (EW) model. The iso-potential (metallic in the terminology of the electrostatic problem) sites are chosen over the lattice according to the percolation problem, giving rise to some metallic islands and some active (not metallic, nor surrounded by a metallic island) area. We see that the dilution of the system by incorporating metallic particles (or equivalently considering the iso-height islands) annihilates the spatial correlations and also the potential fluctuations. Some local and global critical exponents of the problem are reported in this work. The GFF, when simulated on the active area show a cross over between two regimes: small (UV) and large (IR) scales. Importantly, by analyzing the change of exponents (in and out of the critical occupation $p_c$) under changing the system size and the change of the cross-over points, we find two fixed points and propose that GFF$_{p=p_c}$ is unstable towards GFF$_{p=1}$.
\end{abstract}


\maketitle

\section{Introduction}

The coupling of critical statistical models has recently been a subject of intense study~\cite{gefen1980critical,cheraghalizadeh2018self,barat1995statistics,cheraghalizadeh2018gaussian,daryaei2012watersheds,cheraghalizadeh2017mapping,kremer1981self,najafi2016bak,najafi2016water,najafi2016monte}. One way of the coupling of models is to define a model on a host system with internal degrees of freedom whose arrangement is realized by another statistical model~\cite{cheraghalizadeh2017mapping,cheraghalizadeh2018gaussian,cheraghalizadeh2018self}. This type of coupling may help to understand the structure of fixed points of the combined model~\cite{najafi2018coupling}. Among the 2D critical statistical models, the Gaussian free field (GFF) has an especial importance due to its connection to a wide range of statistical models, ranging from free Bosons~\cite{francesco1996conformal}, to stationary state of Edwards-Wilkinson (EW) of growth process of rough surfaces~\cite{stanley2012random}. The coupling of GFF to the other models, is possible via (but not restricted to) the dilution of the host media whose pattern is tuned by the other statistical model, or by distributing some disorders in the regular lattice according to a model that yields the position pattern of the disorders. The example is the Poisson equation due to a white noise charged disorder (which is equivalent to GFF) in the presence of metallic regions whose formation pattern are modeled by the Ising model with an artificial temperature~\cite{cheraghalizadeh2018gaussian}. Such a study is expected to be relevant in understanding of many solid state systems which are doped with the metallic particles with a vast applications ranging from optical devices~\cite{miller2013nonlinear,haglund1993picosecond,shalaev1998nonlinear,cai2005superlens,kravets2010plasmonic,kim2017ultrafast,kim2018dielectric,kim2013nondegenerate,arya1986anderson} and random lasers~\cite{meng2013metal}, to sensor technology~\cite{franke2006metal} and solar cells~\cite{huang2013mitigation,yu2017effects,li2016electrostatically}.\\
Along with the experimental interests, there are also theoretical interests on the various versions of the GFF model. Many condensed matter systems are directly mapped to the Coulomb gases (which, in the zero background charge, corresponds to the GFF model)~\cite{nienhuis1982analytical}. Among them are the XY model~\cite{villain1975j}, the Ashkin-Teller model~\cite{knops1982renormalization}, the $q$-state Potts model~\cite{knops1982renormalization,nienhuis1982analytical}, the antiferromagnetic Potts model~\cite{den1982critical}, the $O(n)$ model, the frustrated Ising models~\cite{nienhuis1982b}, vortices dynamics in superfluids~\cite{kosterlitz1974jm} and the quantum Hall systems via the plasma analogy of the wave function~\cite{girvin1999quantum}. This correspondence is not restricted to the equilibrium phenomena. For example the Edwards-Wilkinson (EW) model of growth process in the stationary state corresponds to GFF~\cite{kondev2000nonlinear,kondev1995geometrical}. On the two-dimensional (2D) regular systems, it is well-known that the GFF belongs to $c=1$ conformal field theory (CFT)~\cite{francesco1996conformal} and also the Coulomb gas with the coupling constant $g=1$~\cite{cardy2005sle}. All of these make the main aim of the present paper, i.e. the effect of environmental disorder on GFF as a long-standing problem in any condensed matter system, very important in both theoretical and experimental sides. In the CFT language, the problem of GFF in the presence of the un-correlated metallic disorder with critical occupation is interpreted as the coupling of $c=1$ CFT with the $c=0$ (critical percolation) CFT, which has poorly been investigated in the literature. Also the structure of presumable fixed points in the off-critical regime is a worthy problem.\\
In the present paper we realize the GFF by considering the Poisson equation in the background of random white-noise charges with normal distribution, and study the effect of metallic regions whose positional configurations are modeled by un-correlated percolation model which is tuned by the occupation probability $p$. We study various local and global statistical observable in terms of $p$ and the behavior of the critical exponents is obtained. Interestingly we observe that there are two regimes with distinct critical behaviors: small (called UV) scales and large (called IR) scales. The IR exponents fit properly to the regular GFF model, i.e. GFF$_{p=1}$, whereas the UV exponents show a new universality class which should be characterized in more details in the community. By analyzing the finite size effects along with the cross-over scale between these behaviors, we propose a fixed point structure for this problem, whose phase space is drawn at the end.\\

The paper has been organized as follows: The SEC.~\ref{Review} is devoted to the some concepts of rough surfaces and the important statistical observables in the problem. In the SEC.~\ref{RoughSurfaces}, we motivate this study and introduce and describe the model. The numerical methods and details are explored in SEC.~\ref{NUMDet}. The results are presented in the section~\ref{results} for local (SEC.~\ref{local}) and global (SEC.\ref{global}) quantities. We end the paper by a discussion and conclusion in SEC.~\ref{conclusion}.

\section{A review on the scale-invariant rough surfaces}\label{Review}
It is necessary to review some features of the scale-invariant 2D random fields and rough surfaces. The methods employed in these systems have wide applications, ranging from the classical~\cite{kondev2000nonlinear,kondev1995geometrical,cheraghalizadeh2018gaussian} to the quantum systems~\cite{Najafi2017Scale,najafi2018scaling,najafi2018percolation,najafi2018interaction}. Let $V(x,y)\equiv V(\mathbf{r})$ be the \textit{height} profile (in this paper the electrostatic potential) of a scale invariant 2D random rough field. The main property of self-affine random fields is their invariance under rescaling \cite{barabasi1995fractal,kirchner2003critical,falconer2004fractal}. The probability distribution function of these fields transform under $\textbf{r}\rightarrow \lambda\textbf{r}$ as follows:
scaling law
\begin{eqnarray}\label{scaleinvariance}
V(\lambda \mathbf{r}) \stackrel{d}{=} \lambda ^{\alpha} V(\mathbf{r}),
\end{eqnarray}
where the parameter $\alpha$ is \textit{roughness} exponent or the \textit{Hurst} exponent and $\lambda$ is a scaling factor and the symbol $\stackrel{d}{=}$ means the equality of the distributions. Let us denote the Fourier transform of $V(\textbf{r})$ by $V(\textbf{q})$. The distribution of a wide variety of random fields characterized by the toughness exponent $\alpha$ is Gaussian with the form
\begin{eqnarray}
P\left\lbrace  V \right\rbrace \sim \exp \left[ -\frac{k}{2} \int _ 0 ^ {q_0} d \mathbf{q}q^{2(1+\alpha)}V_{\mathbf{q}}V_{-\mathbf{q}} \right],
\end{eqnarray}
where $q_0$ is the momentum cut-off which is of the order of the inverse of the lattice constant~\cite{kondev1995geometrical} and $k$ is some constant. The scale invariance, when combined with the translational, rotational and scale invariance, has many interesting consequences. For example the height-correlation function of $V(\mathbf{r})$, $C(r) \equiv \langle \left[ V(\mathbf{r}+\mathbf{r_0})-V(\mathbf{r_0}) \right]^2 \rangle$ is expected to behave like
\begin{eqnarray}\label{height-corr}
C(r) \sim |\mathbf{r}| ^{2\alpha_l},
\end{eqnarray}
where the parameter $\alpha_l$ is called the local roughness exponent \cite{barabasi1995fractal} and $\left\langle \right\rangle$ denotes the ensemble average. The above equation implies that the second moment of $V(\textbf{q})$ scales with $q$ for small values of $q$, i.e. $S(\mathbf{q})\equiv \langle |V(\mathbf{q})|^2\rangle\sim |\mathbf{q}|^{-2(1+\alpha)}$~\cite{falconer2004fractal} which is obtained from the relation \ref{height-corr}. Another measure to classify the scale invariant profile $V(\mathbf{r})$ is the total variance
\begin{eqnarray}\label{total variance}
W(L)\equiv \langle \left[ V(\mathbf{r}) - \bar{V} \right]^2 \rangle _L \sim L^{2\alpha_g}
\end{eqnarray}
where $\bar{V}=\langle V(\mathbf{r}) \rangle_L$, and $\langle \dots \rangle _L$ means that, the average is taken over $\mathbf{r}$ in a box of size $L$. The parameter $\alpha_g$ is the global roughness exponent. Self-affine surfaces are mono-fractals just if $\alpha_g = \alpha_l = \alpha$ \cite{barabasi1995fractal}. \\
The other test for $V(\textbf{r})$ to be Gaussian is that all of its finite-dimensional probability distribution functions are Gaussian \cite{adler1981geometry}. One of the requirements of this is that its distribution is Gaussian:
\begin{eqnarray}
P( V ) \equiv \frac{1}{\sigma\sqrt{2\pi}}e^{-\frac{V^2}{2\sigma^2}},
\end{eqnarray}
where $\sigma$ is the standard deviation. Another quantity whose moments distributions should be Gaussian is the local curvature which is defined (at position $\mathbf{r}$ and at scale $b$) as~\cite{kondev2000nonlinear}
\begin{eqnarray}\label{local curvature}
C_b(\mathbf{r}) = \sum_{m=1}^M \left[ V(\mathbf{r}+ b\mathbf{e}_m) - V(\mathbf{r}) \right],
\end{eqnarray}
in which the offset directions $\left\lbrace\mathbf{e}_1,\dots,\mathbf{e}_M\right\rbrace$ are a fixed set of vectors whose sum is zero, i.e. $\sum_{m=1}^M \mathbf{e}_m =0$. If the rough surface is Gaussian, then the distribution of the local curvature $P (C_b)$ is Gaussian and the first and all the other odd moments of $C_b$ manifestly vanish since the random field has up/down symmetry $V(\mathbf{r})\longleftrightarrow -V(\mathbf{r})$. Additionally, for Gaussian random fields we have:
\begin{eqnarray}\label{fourth moment}
\frac{\langle C_b^4 \rangle }{\langle C_b^2 \rangle ^2} = 3.
\end{eqnarray}
This relation is an important test for the Gaussian/non-Gaussian character of a random field. \\

All of the analysis presented above are in terms of local variable $V(\textbf{r})$. There is however a non-local view of point in such problems, i.e. the iso-height lines of the profile $V(\mathbf{r})$ at the level set $V(\mathbf{r}) = V_0$ which also show the scaling properties. When we cut the self-affine surface $V(x,y)$ some non-intersecting loops result which come in many shapes and sizes \cite{kondev1995geometrical,kondev2000nonlinear}. We choose $10$ different $V_0$ between maximum and minimum potentials and a \textit{contour loop ensemble} (CLE) is obtained. These geometrical objects are scale invariant and show various power-law behaviors, e.g., their size distribution is characterized by a few power law relations and scaling exponents. The scaling theory of CLEs of self-affine Gaussian fields was introduced in Ref. \cite{kondev1995geometrical} and developed in Ref. \cite{kondev2000nonlinear}. In the following we introduce the various functions and relation, firstly introduced in Ref. \cite{kondev2000nonlinear}. The most important local quantities are $\alpha_l$ and $\alpha_g$. For the non-local quantities the exponent of the distribution functions of loop lengths $l$ ($P(l)$) and the gyration radius of loops $r$ ($P(r)$) are of especial importance. In addition the contour loop ensemble can be characterized through the loop correlation function $G(\mathbf{r})=G(r)$ ($r\equiv |\textbf{r}|$) which is the probability measure of how likely the two points separated by the distance $r$ lie on the same contour. For large $r$s this function scales with $r$ as
\begin{eqnarray}\label{loop correlation function}
G(r) \sim \frac{1}{r^{2x_l}},
\end{eqnarray}
where $x_l$ is the loop correlation exponent. It is believed that the exponent $x_l$ is superuniversal, i.e. for all the known mono-fractal Gaussian random fields in two dimensions this exponent is equal to $\frac{1}{2}$~\cite{kondev1995geometrical,kondev2000nonlinear}. \\
Now consider the probability distribution $P(l,r)$ which is the measure of having contours with length $(l, l + dl)$ and radius $(r, r + dr)$. For the scale invariant CLE, $P(l,r)$ is hypothesized to behave like~\cite{kondev1995geometrical}:
\begin{eqnarray}
P(l,r) \sim l^{-\tau_l -1/D_f} g(l/r^{D_f}),
\end{eqnarray}
where $g$ is a scaling function and the exponents $D_f$ and $\tau_l$ are the fractal dimension and the length distribution exponent, respectively. One also can define the fractal dimension of the loops by the relation $\langle l \rangle \sim r^{\gamma_{lr}}$. By the following straightforward calculation
\begin{eqnarray}\label{loop fractal dimesion}
\langle l \rangle \equiv \frac{\int _0^\infty lP(l,r) dl}{\int _0^\infty P(l,r) dl} \sim r^{D_f},
\end{eqnarray}
we see that $\gamma_{lr}=D_f$. Note also that the probability distribution of contour lengths $P(l)$ is obtained using the relation $P(l) \equiv \int_0^\infty P(l,r)dr \sim l^{-\tau_l}$. It is shown that there are the important scaling relations between the scaling exponents $\alpha$, $D_f$, $\tau_l$ and $x_l$ as follows~\cite{kondev2000nonlinear}:
\begin{eqnarray}\label{hyper1}
D_f(\tau_l -1) = 2-\alpha,
\end{eqnarray}
and
\begin{eqnarray}\label{hyper2}
D_f(\tau_l-3) = 2x_l -2.
\end{eqnarray}\\

In the general theory of critical phenomena, each system in the critical state shows some power-law behaviors for the local and geometrical quantities, i.e. $P(x)\sim x^{-\tau_x}$ ($x=$ the local and geometrical statistical quantities). The estimation of these exponents is a challenging problem, for which a detailed finite-size analysis is required. For mono-fractal finite systems, the finite-size scaling (FSS) theory predicts that~\cite{goldenfeld1992lectures}:
\begin{equation}
P_x(x,L)=L^{-\beta_x}g_x(xL^{-\nu_x}),
\label{eq:FSS}
\end{equation}
in which $g_x$ is a universal function and $\beta_x$ and $\nu_x$ are some exponents that are related by $\tau_x=\frac{\beta_x}{\nu_x}$. For multi-fractal systems, this prediction does not work. For example, for a system with two distinct critical regions (UV and IR regions in our main problem), one expects some cross-over point $x^*$ which connects these two regions~\cite{najafi2012avalanche,najafi2016bak,najafi2016water}. For determining these points, we have followed the method presented in~\cite{najafi2018statistical}, in which the slope of each part of the graph is obtained (in the log-log plot) and the cross-over point is obtained as the point in which the linear fits meet each other. The exact determination of these points is not simple in simulations, since when the exponents of the two regions are close to each other, the statistical error bar for $x^*$ becomes large~\cite{najafi2018statistical}.\\
Finally we note that there is a hyper-scaling relation between the $\tau$ exponents and the fractal dimensions $\gamma_{x,y}$, which are defined by the relation $x\sim y^{\gamma_{x,y}}$, namely:
\begin{equation}
\gamma_{x,y}=\frac{\tau_y-1}{\tau_x-1}.
\end{equation}
This relation is valid only when the conditional probability function $p(x|y)$ is a function with a very narrow peak for both $x$ and $y$ variables.

\section{The construction of the problem}\label{RoughSurfaces}
\label{sec:model}

In this section we construct the main idea of the present paper. The Gaussian free field (GFF) is a very important model to which many models are mapped, ranging from free Boson filed, to Edwards-Wilkinson (EW) model of surface growth process. A realization of GFF is the Poisson equation in the background of white-noise charge disorders, which itself is mapped to EW model in the stationary state. If other kinds of disorder is present in the system (that is the case for any condensed matter system), the problem of the statistics of the electric potential becomes complicated, since the GFF is coupled to the other model which realizes the disorder. There are many experimental~\cite{miller2013nonlinear,haglund1993picosecond,shalaev1998nonlinear,cai2005superlens,kravets2010plasmonic,kim2017ultrafast,kim2018dielectric,kim2013nondegenerate,meng2013metal,subramanian2001semiconductor,huang2013mitigation,yu2017effects,li2016electrostatically} and theoretical~\cite{cheraghalizadeh2018gaussian} motivations to consider the metallic disorders in dielectric media. If the mentioned disorders are some metallic particles randomly distributed over the sample (which is the subject of the present paper), then the problem is simply finding the solution of the Poisson equation with some additional boundary conditions imposed by metallic regions. The configuration of the position of metallic particles is naturally random and may be modeled by some well-understood models, like the percolation theory (for uncorrelated metallic disorders). In this case, noting that in the absence of iso-potential islands the system corresponds to Gaussian free field (GFF), we can imagine of this problem as \textit{the coupling of the GFF with the percolation theory as a model of the position pattern of the metallic islands}. It is the easiest way of configuring metallic particles in the media, although the spatial pattern of connectedness of metallic particles is generally complex and many internal degrees of freedom play role in the problem, e.g. the cohesive energy, the particle sizes and the effect of the media around. When the host configuration is made, one can simulate the dynamical (GFF) model, assuming that the metallic islands are quenched. Some other examples of coupling of statistical models can be seen in~\cite{najafi2018coupling,najafi2016monte}.\\
For the purposes mentioned above, the system is meshed by cells each of which can have one of the two states: empty or occupied by a metallic particle (which we call metallic site). The metric space is therefore tuned by the occupation probability $p$ which is the probability that a site is empty (not occupied by a metallic particle). Then the Poisson equation is solved in the background of the metallic islands for the white-noise random charges in the non-metallic (active) area. \\
Before describing the problem in this type of media, let us first briefly introduce the standard method of generating GFFs. As mentioned above, the EW model in the stationary state becomes GFF which is generated by the following equation for the height field $V(\textbf{r})$:
\begin{equation}
\partial_t V(\vec{r},t)=\nabla^2V(\vec{r},t)+\eta(\vec{r},t),
\label{Eq:EW}
\end{equation}
in which $\eta(\vec{r},t)$ is a space-time white noise with the properties $\left\langle \eta(\vec{r},t)\right\rangle = 0 $ and $\left\langle \eta(\vec{r},t)\eta(\vec{r}',t')\right\rangle = \zeta \delta^3(\vec{r}-\vec{r}')\delta(t-t') $ and $\zeta$ is the strength of the noise. $V(\vec{r},t)$ can be served as the electrostatic potential in our paper, once it becomes $t$-independent (the stationary state of EW, in which $\partial_t V=0$), acquiring the following form (with the dielectric constant $\epsilon\equiv 1$):
\begin{equation}
\nabla^2V(\vec{r})=-\rho(\vec{r}),
\label{Eq:Poisson}
\end{equation}
where $\rho(\vec{r})$ is the spatial white noise with the normal distribution and the properties $\left\langle \rho(\vec{r})\right\rangle = 0 $ and $\left\langle \rho(\vec{r})\rho(\vec{r}')\right\rangle = (n_ia)^2 \delta^3(\vec{r}-\vec{r}')$, $n_i$ is the total density of Coulomb disorder, $a$ is the lattice constant. It is well-known that this model in the scaling limit belongs is described by Gaussian distribution function (GFF) which is $c=1$ conformal filed theory~\cite{francesco1996conformal}. It is also known that the contour lines of this model are described by the Schramm-Loewner evolution (SLE) theory with the diffusivity parameter $\kappa=4$~\cite{cardy2005sle}, which is understood in terms of the general CFT/SLE correspondence with the relation $c=(6-\kappa)(3\kappa-8)/(2\kappa)$. The fractal dimension of the contour loops $D_f^{\text{GFF}}=\frac{3}{2}$ which is also compatible with the relation $D_f=1+\frac{\kappa}{8}$. \\
Now let us consider the problem of GFF in the background of metallic islands. The effect of these islands is that, over them the potential is constant, i.e. the Neumann type boundary conditions. This cause the contour lines of potential are deformed and also the fluctuations of the potential are changed (as becomes clear in the following sections). To determine the shape of these islands, the percolation theory is defined on the $L\times L$ square lattice with the occupation probability $p$ and the connected clusters composed of active (non-metallic) sites are identified. The \textit{active space} is defined as the set of sites that are un-occupied and also are not completely surrounded by a metallic island, i.e. there are some free paths from the site to infinity (or system boundaries). The random charged impurities are put on the sites in the active space and the Poisson equation is solved with the imposed boundary condition, that is free in our paper. \\
This problem belongs also to the context of the critical phenomena on the fractal systems. This concept was mainly begun by the work of Gefen \textit{et. al.}~\cite{gefen1980critical} in which it was claimed that the critical behavior of the models is tuned by the detail of the topological quantities of the fractal lattice. The cluster fractal dimension, the order of ramification and the connectivity are some examples of these quantities~\cite{gefen1980critical}. This concept can be extended to dilute systems that are fractal in some limits~\cite{cheraghalizadeh2017mapping,najafi2016bak,najafi2016monte,najafi2016water,najafi2018coupling}. There are also some experimental motivations for such studies. The examples is the magnetic materials in the porous media \cite{kose2009label,kikura2004thermal,matsuzaki2004real,philip2007enhancement,kim2008magnetic,keng2009colloidal,kikura2007cluster,najafi2016monte}. In this paper the critical percolation model ($p=p_c$ in which $p_c$ is the critical occupation for the percolation model above which there is almost surely one percolated cluster, i.e. a cluster of same type which connects two apposite boundaries) plays the role of the fractal lattice on which the GFF is considered. For the case $p=1$ which is a regular lattice, one retrieves the results of the ordinary $c=1$ CFT and also SLE$_{4}$.\\ 
The off-critical occupations, i.e. $p>p_c$, are also very important, especially in the close vicinity of the $p_c$, which help to determine the universality class of the model. In this regime, some critical exponent may be obtained. Also in many cases, the off-criticality parameter (here $\epsilon_0\equiv 1-p$) drive the original critical model (here GFF$_{p=1}$, i.e. regular GFF) towards some other fixed point. The relevance of irrelevance of this perturbation should be deduced from some numerical evidences, which is a part of aims of this paper.

\subsection{Numerical methods}\label{NUMDet}

\begin{figure*}
	\centering
	\begin{subfigure}{0.40\textwidth}\includegraphics[width=\textwidth]{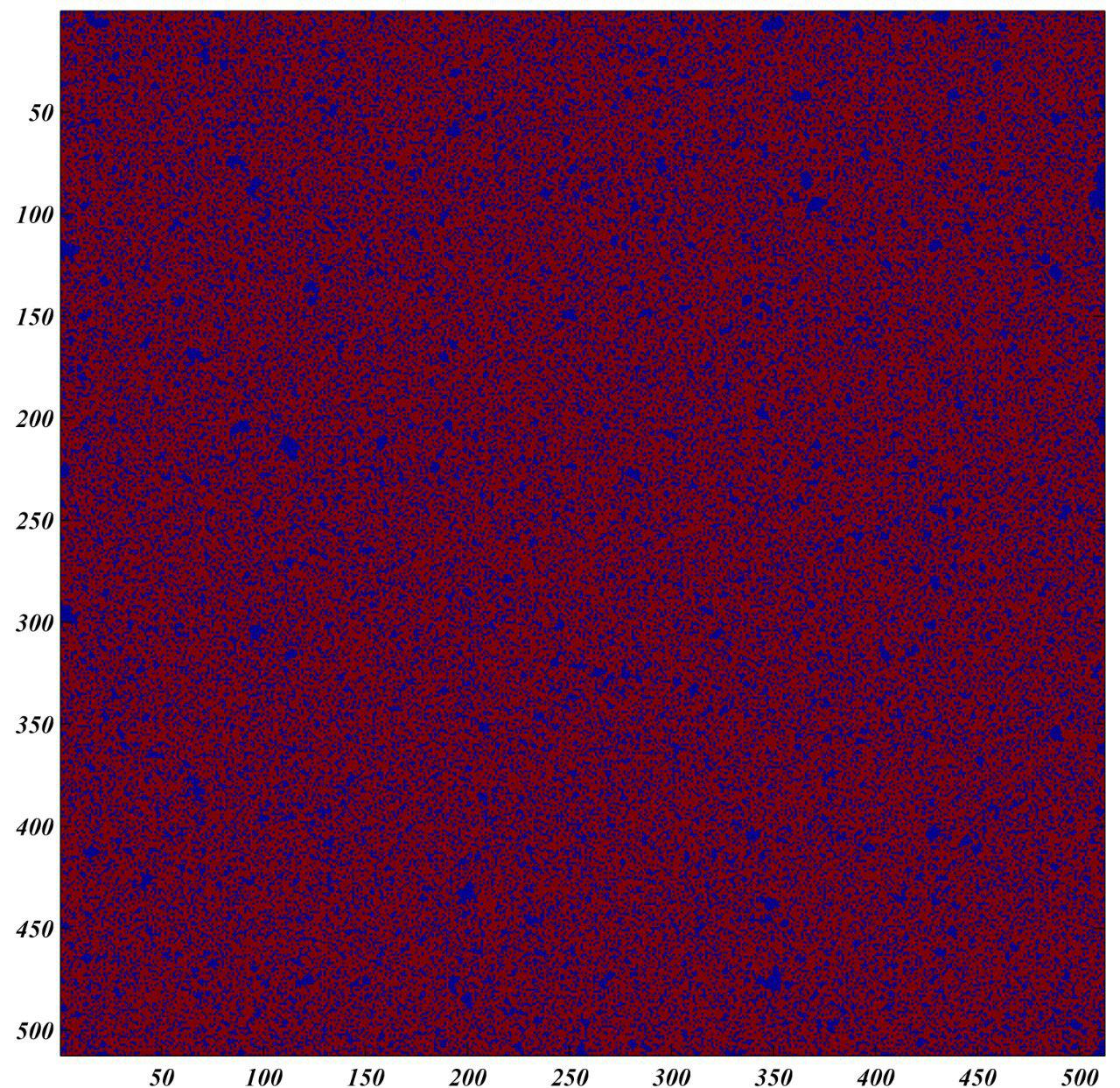}
		\caption{}
		\label{fig:pGreaterThanPcSample}
	\end{subfigure}
	\begin{subfigure}{0.45\textwidth}\includegraphics[width=\textwidth]{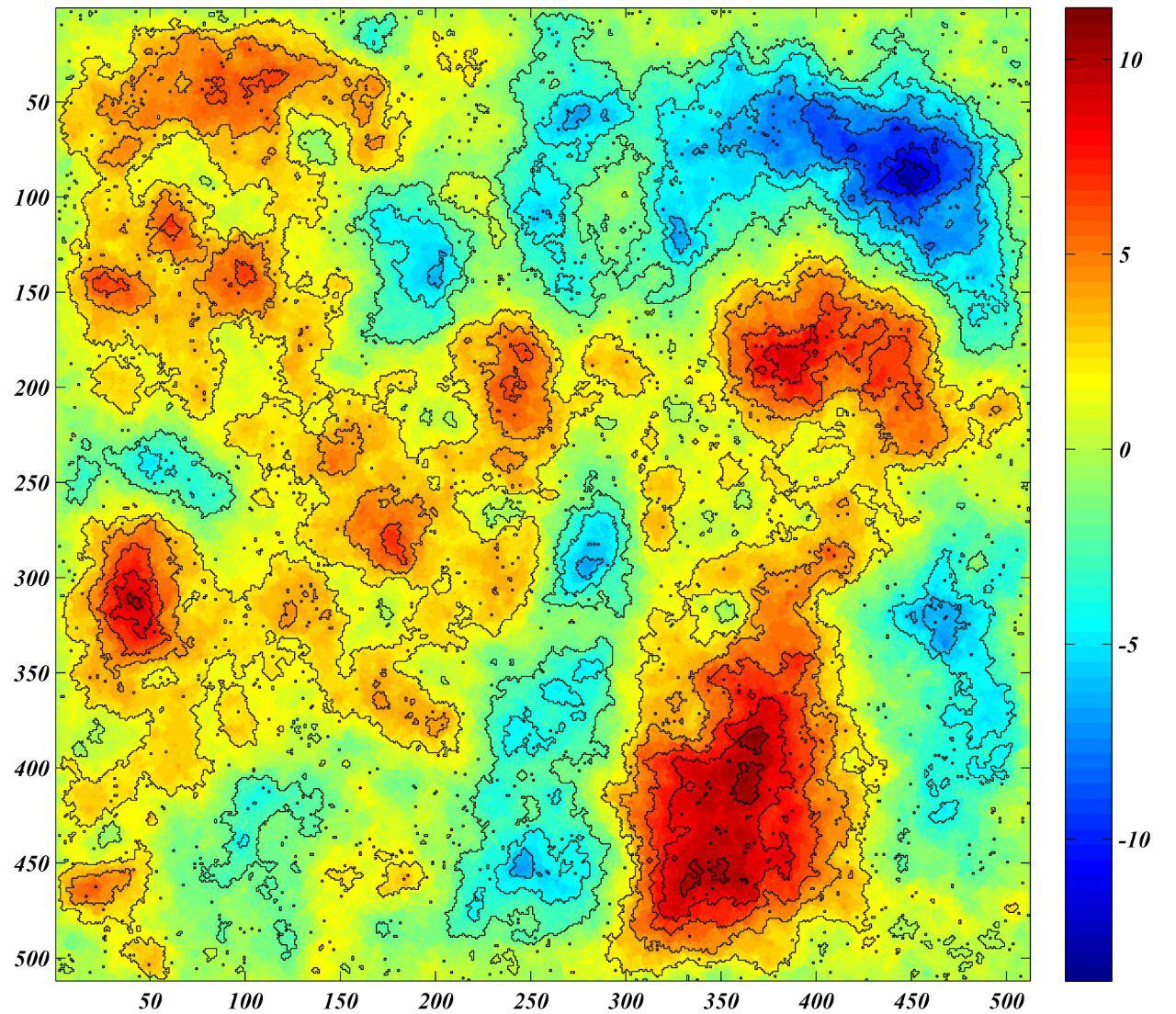}
		\caption{}
		\label{fig:pGreaterThanPcSolution}
	\end{subfigure}
	\begin{subfigure}{0.40\textwidth}\includegraphics[width=\textwidth]{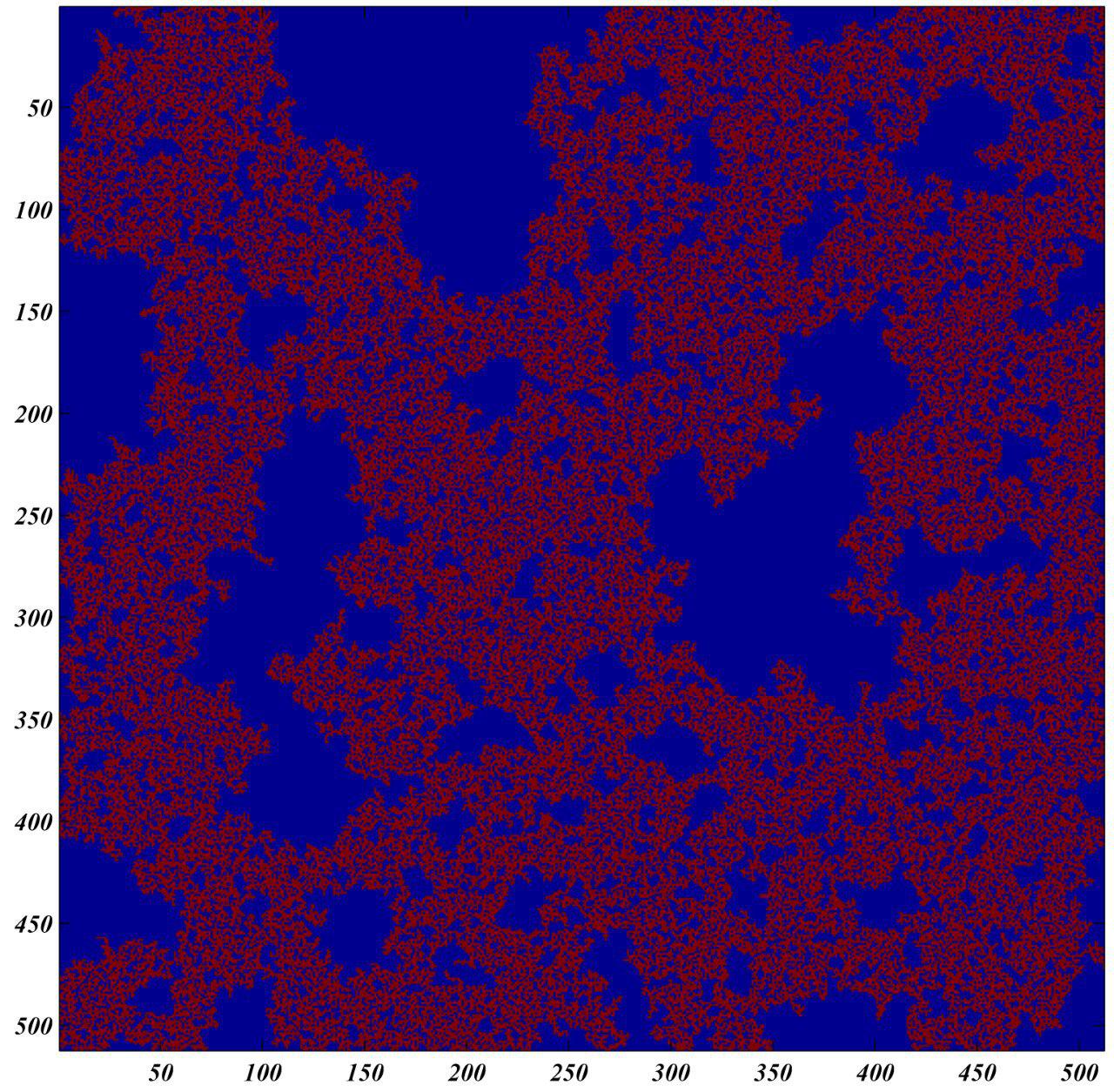}
		\caption{}
		\label{fig:pcSample2}
	\end{subfigure}
	\begin{subfigure}{0.45\textwidth}\includegraphics[width=\textwidth]{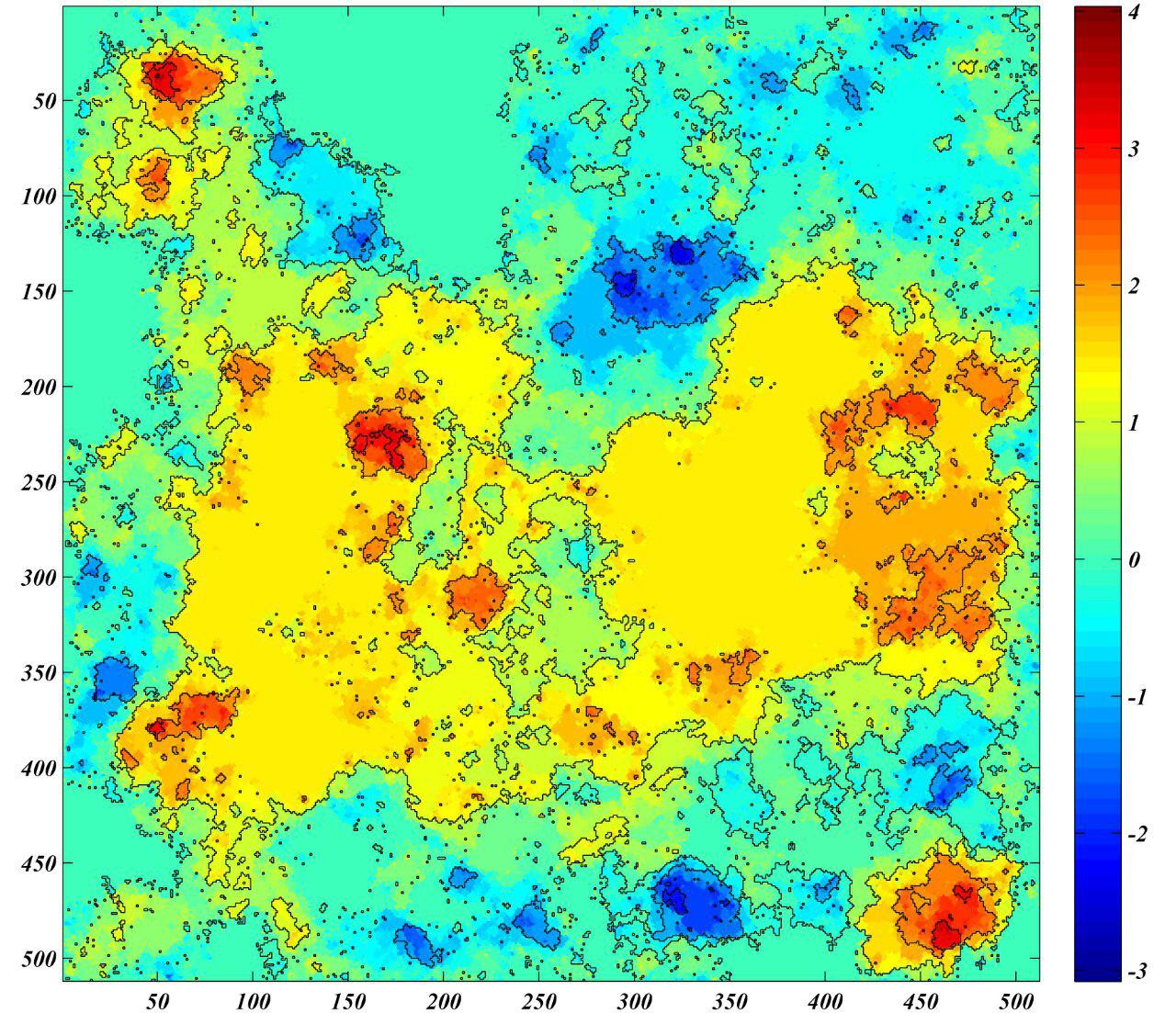}
		\caption{}
		\label{fig:pcSolution2}
	\end{subfigure}
	\caption{(Color online) The (a) sample and (b) the corresponding electrostatic potential for $p=0.70>p_c$. The (c) sample and (d) the corresponding electrostatic potential for $p=p_c$.}
	\label{fig:samples}
\end{figure*}

As explained in the previous subsection, we consider $L\times L$ square lattice and put some metallic particles in some random sites over the lattice in such a way that the mentioned site is completely covered by the metallic particle. Each site is occupied by a metallic particle with the probability $1-p$ and is un-occupied (active) with the probability $p$. When an electrostatic potential is obtained, we extract the contour lines by $10$ different cuts with the same spacing between maximum and minimum values. We have run the program for lattice sizes $L=256, 512$ and $1024$ to control the finite size effects. It is notable that for each $L=1024$ sample (for a given $p$) about $\sim 10^3$ loops were obtained. This means that for each $p$ and $L$, something like $10^8$ loops were generated. The Hoshen-Kopelman~\cite{hoshen1976percolation} algorithm has been employed for identifying the clusters in the lattice. Figure \ref{fig:samples} shows samples and their corresponding potentials for two cases: $p=p_c$ and $p>p_c$. We see that the metallic islands become larger and self-similar as $p$ approaches $p_c$. The contour lines for $10$ different cuts have also been shown.

\section{Results}\label{results}

We have simulated the system for some $p\geq p_c=0.5927$. The samples, along with the electrostatic solutions have been shown in Fig.~\ref{fig:samples}. The blue area in the percolation samples is the metallic area or is the area which is surrounded by metallic particles, and is therefore iso-potential region. For all occupation probabilities ($p$), the system show critical behaviors and the random potential pattern $V(\textbf{r})$ is self-similar and scale invariant. However these critical behaviors are not the same for all scales, i.e. for all statistical observables we have observed two distinct regions with their own critical properties. One of them governs the properties of the model in small scales, and the other controls the large scale behaviors. We call the former as the \textit{UV properties} and the latter as the \textit{IR properties}. In the following sections we report the critical properties of the model in each region. Our study contains two separate parts: the local observables and the global ones. 

\subsection{local properties}\label{local}

\begin{figure*}
	\centering
	\begin{subfigure}{0.49\textwidth}\includegraphics[width=\textwidth]{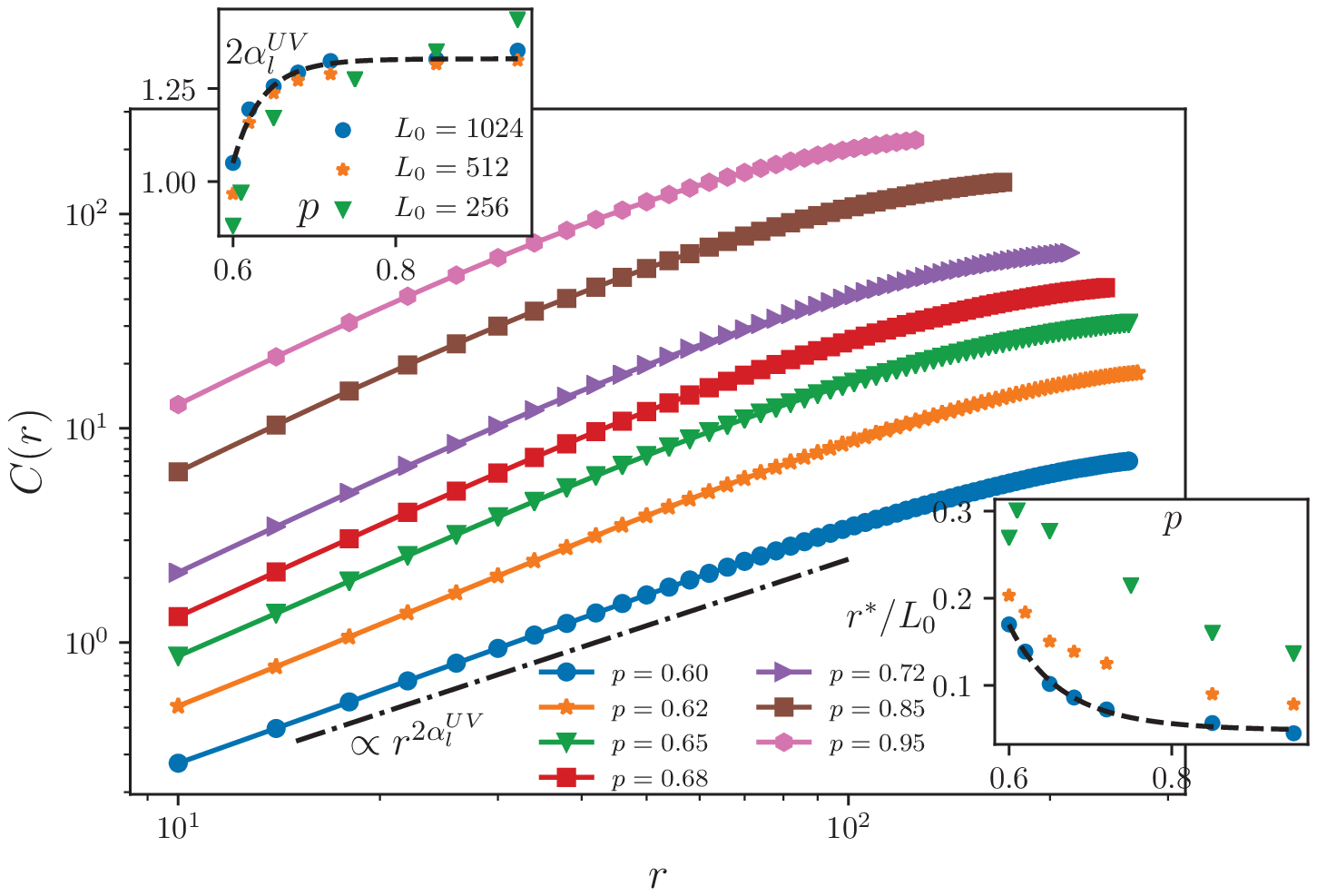}
		\caption{}
		\label{fig:C-r}
	\end{subfigure}
	\begin{subfigure}{0.49\textwidth}\includegraphics[width=\textwidth]{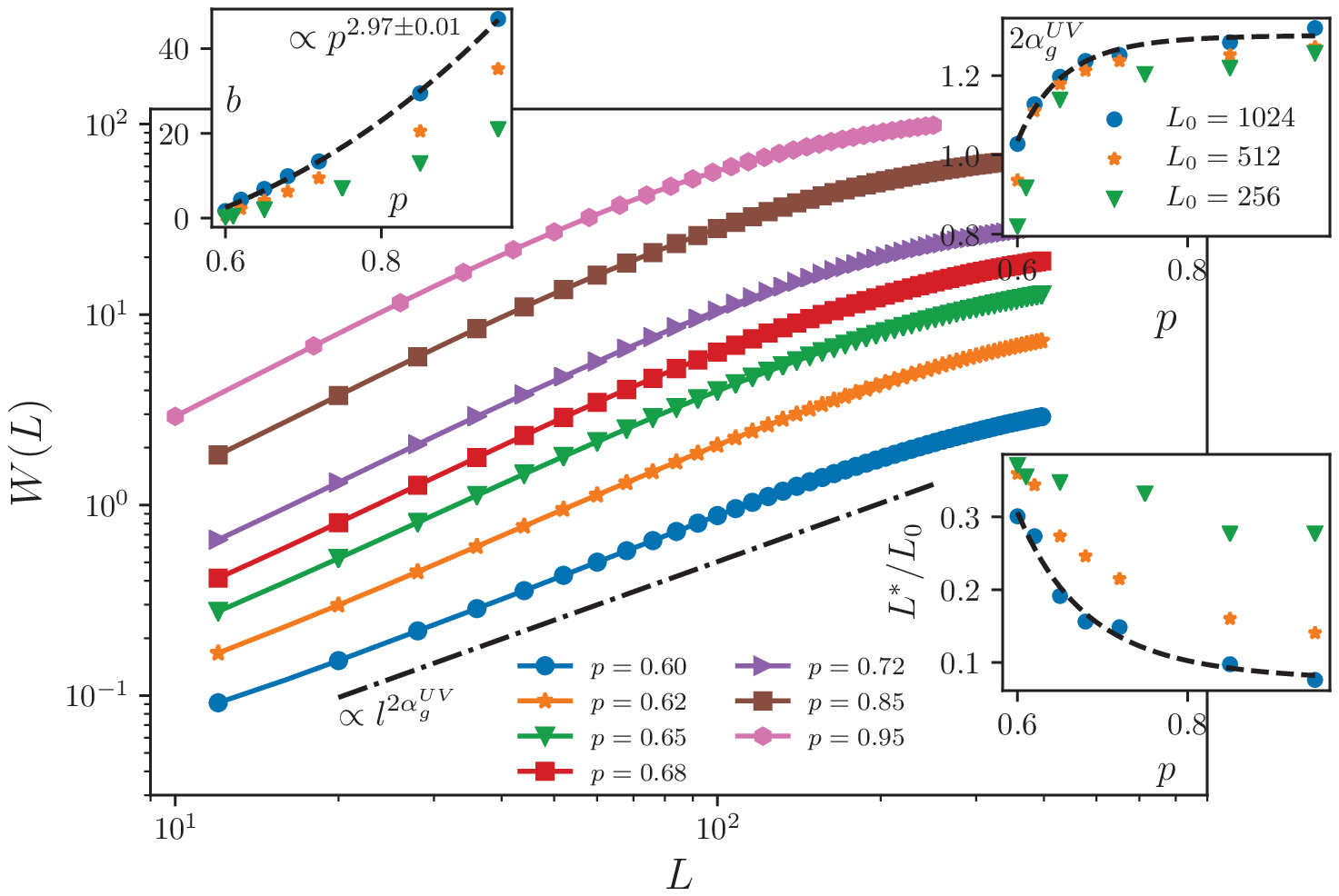}
		\caption{}
		\label{fig:W-l}
	\end{subfigure}
	\begin{subfigure}{0.49\textwidth}\includegraphics[width=\textwidth]{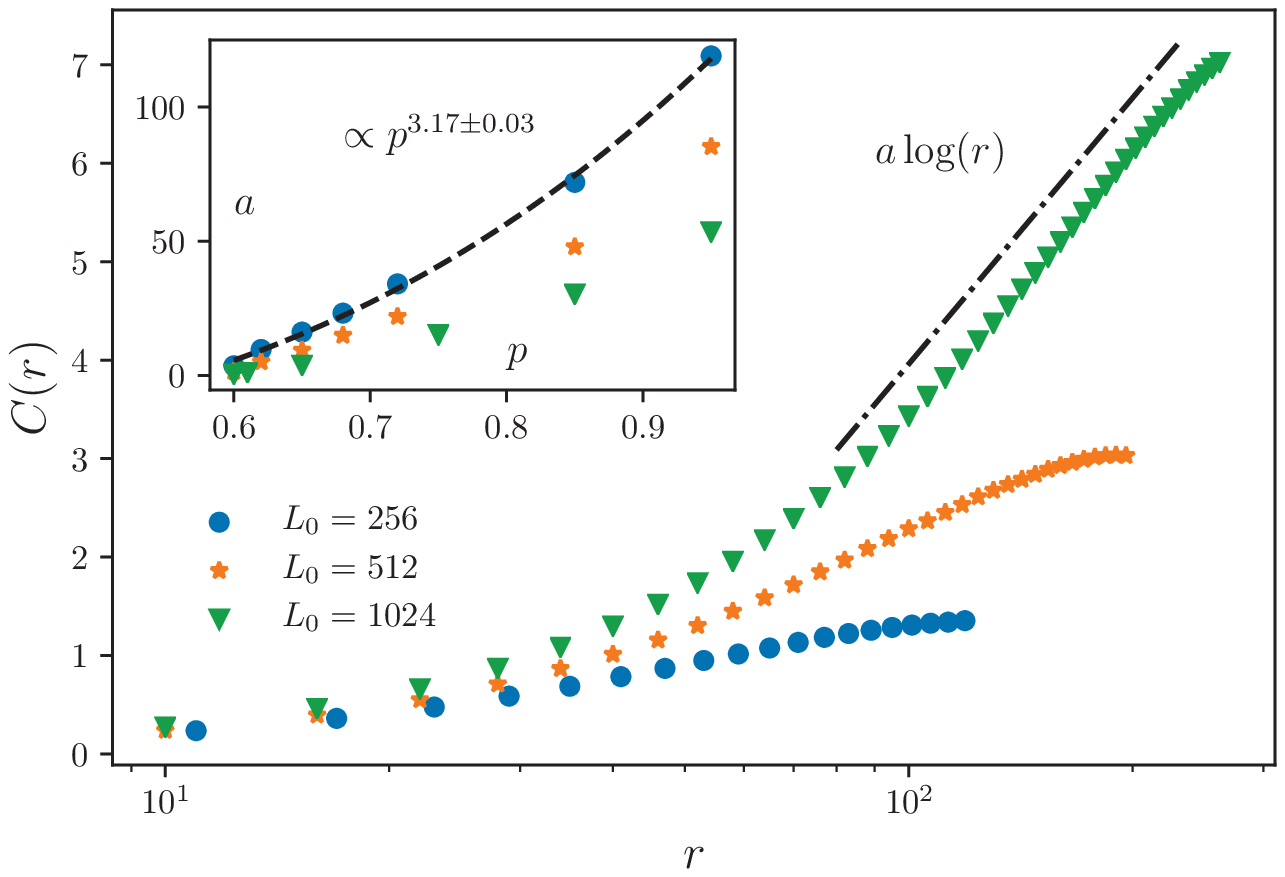}
		\caption{}
		\label{fig:C-r-L}
	\end{subfigure}
	\begin{subfigure}{0.49\textwidth}\includegraphics[width=\textwidth]{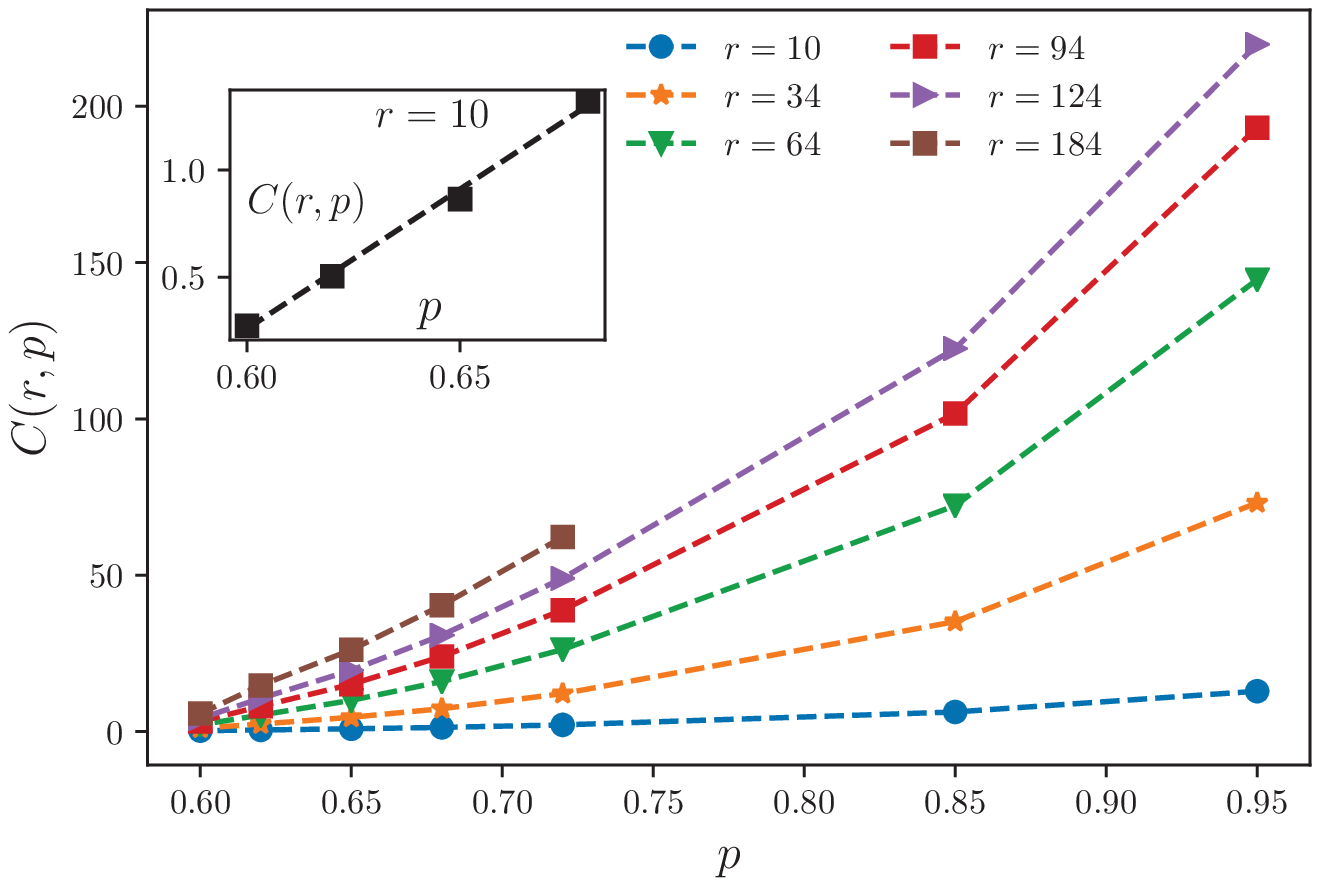}
		\caption{}
		\label{fig:C_r_p}
	\end{subfigure}
	\begin{subfigure}{0.49\textwidth}\includegraphics[width=\textwidth]{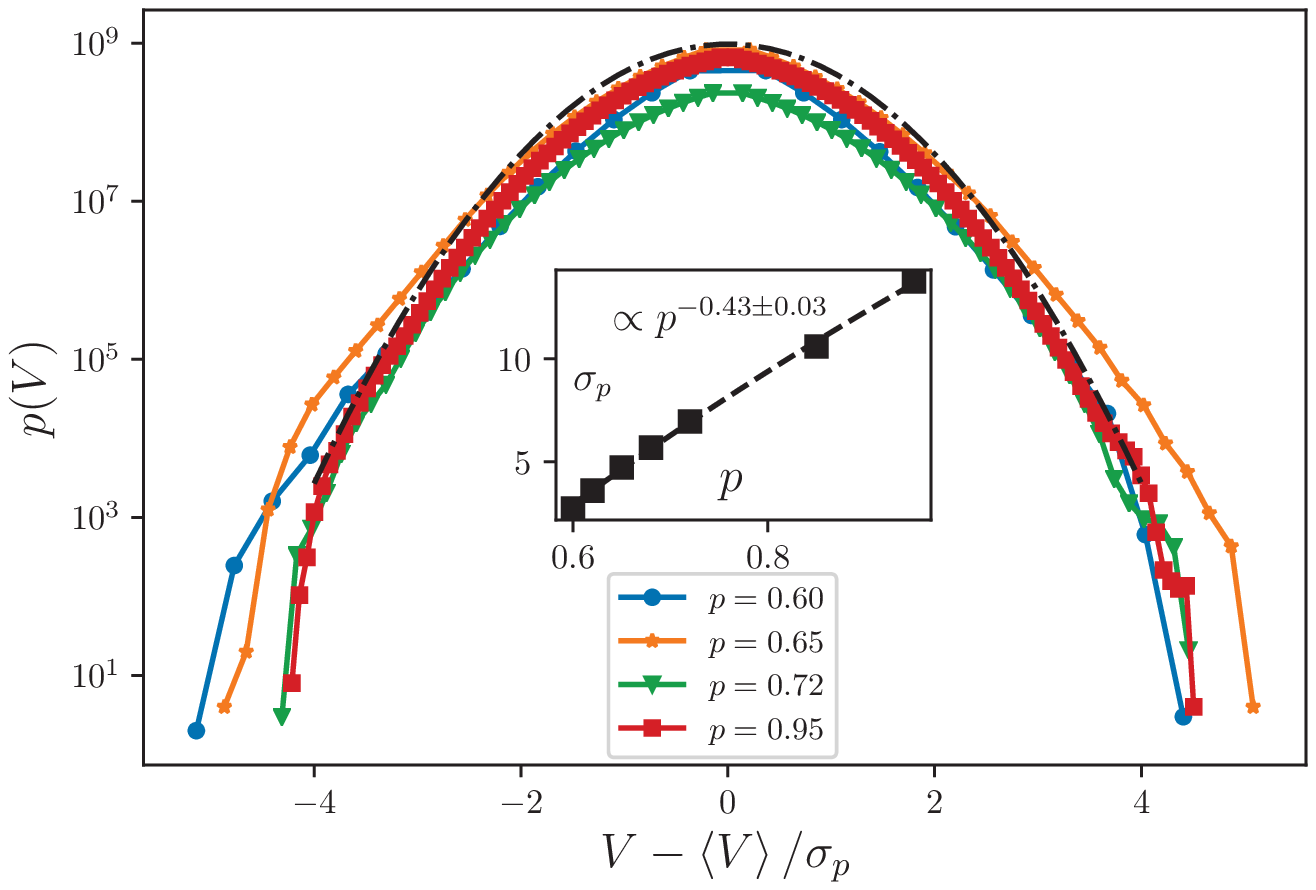}
		\caption{}
		\label{fig:p-h}
	\end{subfigure}
	\begin{subfigure}{0.49\textwidth}\includegraphics[width=\textwidth]{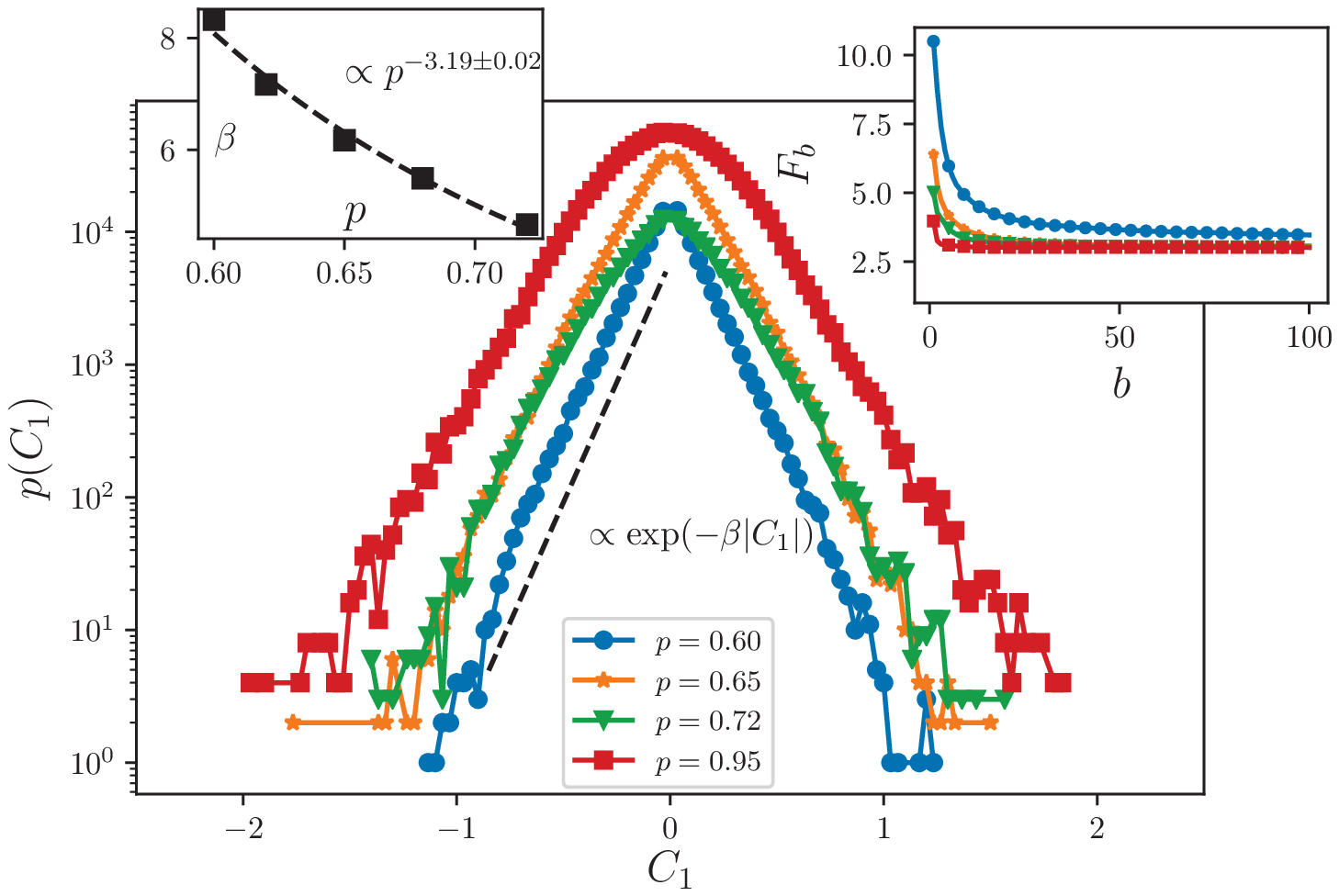}
		\caption{}
		\label{fig:Cb}
	\end{subfigure}
	\caption{(Color online) The behavior of (a) $C(r)$ and (b) $W(l)$ with the power-law behavior in the small scales and the logarithmic behavior in the large scales. The upper insets show the exponents $\alpha_l^{\text{UV}}$ and $\alpha_g^{\text{UV}}$ for the small scales. The lower insets are the cross-over points $r^*/L_o$ and $L^*/L_0$ in terms of $p$ for various system sizes. The proportionality constants $a(p)$ (for $C(r)$) and $b(p)$ (for $W(L)$) also have been shown. (c) The finite size dependence of $C(r)$ at $p=0.6$. (d) The $p$ dependence of $C(r,p)$. (e) The distribution of the random potential $P(V)$ with respect to $\left( V-\left\langle V\right\rangle \right) /\sigma_p$ for various amounts of $p$ for $L=1024$. Note that $\left\langle V\right\rangle=0$ for all $p$s. Inset: the variance $\sigma_p$ in terms of $p$. (f) The distribution of the curvature of the random potential $P(C_1)$ for various amounts of $p$. The fit is $\exp \left[ -\beta\left|C_1\right|\right] $. Left inset: $\beta$ in terms of $p$. Right inset: $F_b\equiv \frac{\left\langle C_1^4\right\rangle }{\left\langle C_1^2\right\rangle^2}$ in terms of $p$.}
	\label{fig:Off-Tc}
\end{figure*}

The scale-invariant rough surfaces have especial local and global properties, that were reviewed in SEC~\ref{RoughSurfaces}. The most important local exponent in a scale-invariant rough surface is the $\alpha$ exponent defined in Eqs~\ref{height-corr} and \ref{total variance} and generally via the relation~\ref{scaleinvariance}. These quantities have been shown in Figs~\ref{fig:C-r} and~\ref{fig:W-l}. In both graphs, two regions are distinguishable. For small scales (small $r$s in Fig.~\ref{fig:C-r} and small $L$s in Fig.~\ref{fig:W-l}) the behavior is power-law, whereas for the large scales the behavior is logarithmic (with $\alpha_l^{\text{IR}}=\alpha_g^{\text{IR}}=0$). Since the logarithmic behavior is characteristics of the GFF in the regular systems, we conclude that the large-scale (IR) properties of the model are described by this model, i.e. the GFF in the regular systems. This is confirmed by calculating of the geometrical exponents, to be reported in the following section. On the other hand the UV region is characterized by a non-zero $\alpha_l^{\text{UV}}$ and $\alpha_g^{\text{UV}}$ whose dependence on $p$ have been shown in the upper-right insets of Figs~\ref{fig:C-r} and~\ref{fig:W-l}. Our observations in this paper support the fact that there are two points in the phase space with robust exponents: $p=p_c$ (named as UV fixed point, i.e. GFF$_{p=p_c}$) and $p=1$ (named as IR fixed point, i.e. GFF$_{p=p_c}$), between which a cross-over occurs. This cross-over takes place in some point, named as $r^*$ in Fig.~\ref{fig:C-r} and as $L^*$ in Fig.~\ref{fig:W-l}. The determination of the $\alpha_l^{\text{UV}}$ and $\alpha_g^{\text{UV}}$ exponent (and all other exponents) needs the determination of $r^*$ and $L^*$ which is determined by the linear fit of the log-log plot and is defined as the point at which the $R^2$ of the linear fit becomes lower than a threshold, i.e. $0.9$ in this paper. We see that $r^*$ and $L^*$ decrease with $p$ (in a power-law fashion) and vanish for large enough $p$s, where the logarithmic dependence (assigned to the IR region) dominates the graphs. An interesting observation is that $r^*_p(L_0)/L_0$ and $L^*_p(L_0)/L_0$ are decreasing functions of both $L_0$ and $p$ (see the lower insets of Figs~\ref{fig:C-r} and~\ref{fig:W-l}). $r^*_p(L_0)/L_0$ and $L^*_p(L_0)/L_0$ are the parameters that separate IR and UV behaviors and the fact that their dependence on $p$ and $L_0$ are qualitatively the same shows that large (small) scales and large (small) $p$s favor the same regime, i.e. IR (UV) regime. In the other words, having in mind that for larger system sizes (larger $L_0$) the IR properties of any system are more seen and therefore the IR regime dominates the corresponding graphs, we interpret the above observation (decreasing of $r^*_{L_0}(p)/L_0$ and $L^*_{L_0}(p)/L_0$ in terms of $L_0$ and $p$) to mean that \textit{GFF$_{p=1}$ is the IR fixed point towards which GFF$_{p=p_c}$ is unstable}.  The fact that the region of UV (power-law) behaviors shrinks to zero as $p$ increases, supports this hypothesis. Such a behavior is regularly seen in the other statistical observable, as we will see in the following section. \\
Our numerical results show that $\alpha_l^{\text{UV}}\approx \alpha_g^{\text{UV}}=0.5\pm 0.1$. In the IR region however we have $C(r)=a\log r$ and $W(L)=b\log L$ (corresponding to $\alpha_l^{\text{IR}}=\alpha_g^{\text{IR}}=0$) with some proportionality constants $a$ and $b$ which have been reported in the insets. The finite size dependence of $C(r)$ has been shown in Fig.~\ref{fig:C-r-L} in which the logarithmic behavior has been explicitly shown in a semi-logarithmic graph. The decreasing of $a$ and $b$ shows that \textit{the site-dilution of the system decreases the spatial correlations of the potential field and also the roughness of the system in the IR regime}. It has explicitly displayed in Fig.~\ref{fig:C_r_p} in which $C(r,p)$ has been sketched in terms of $p$ for various rates of $r$, according to which we see that $C(r,p)$ decreases with decreasing $p$. $C(r,p)$ changes linearly with $p-p_c$ which has been shown in the inset of Fig.~\ref{fig:C_r_p} for $r=10$, and is correct for all $r$s. According to these observations, we propose that this function has the following form in the vicinity of $p=p_c$:

\begin{equation}
C(r,p)\sim (p-p_c)\times \begin{cases} \log r & \text{ for large} \ r \\
r^{\alpha_l^{\text{UV}}} & \text{ for small} \ r
\end{cases}
\end{equation}

The graphs show deviations from this relation for larger $p$s. Note also that $C(r,p)$ becomes vanishingly small as $p\rightarrow p_c$, which confirms that the dilution of the system suppresses the spatial correlations.

One of the important issues in the random field surfaces is the Gaussian/non-Gaussian properties. The question whether the distribution of a random field is Gaussian can be addressed directly by calculating the distribution function of the field itself (here $P(V)$), and the corresponding curvature field $P(C_b)$ as explained in the previous section. The fact that these functions are Gaussian is a necessary, but not sufficient condition for Gaussian random fields. We have shown these quantities in Figs.~\ref{fig:p-h} and~\ref{fig:Cb}. We see that $P(V)$ preserves the Gaussian form for all $p$s, however the width of the distribution ($\sigma_p$) changes. From the inset, it is inferred that $\sigma_p\sim p^{-0.43\pm 0.03}$ for $L_0=1024$. On the other hand, the fact that $P(C_1)\sim \exp\left( -\beta |C_1|\right)$, ($\beta\sim p^{-3.19\pm 0.02}$ in the close vicinity of $p=p_c$), reveals that for $p\ne 1$ we have non-Gaussian surface. To test this more precisely, we have calculated $F_b\equiv \frac{\left\langle C_b^4\right\rangle }{\left\langle C_b^2\right\rangle^2}$ (which should be equal to $3$ for a Gaussian random field) in terms of $b$ in the inset of Fig.\ref{fig:Cb}. We see that $F_b(p=0.95)$, after some changes for small $b$s, is fixed to $3$ for larger $b$s, whereas the final values for the other $p$s are different, which confirm that the surface becomes non-Gaussian for smaller $p$s, especially at $p=p_c$. Therefore it is important to note that \textit{the GFF$_{p=p_c}$ fixed point is not a Gaussian field, and therefore the Kondev hyperscaling relations are not hold}.\\
The total exponents of the local observables have been gathered in TABLE~\ref{tab:local-exponents}. 

\begin{table}
	\begin{tabular}{c|c|c}
		\hline Exponent & Definition & value \\
		\hline $\alpha_l^{\text{UV}}$ & $C^{\text{UV}}(r)\sim r^{\alpha_l^{\text{UV}}}$ & $0.5\pm 0.1$ \\
		\hline $\alpha_g^{\text{UV}}$ & $W^{\text{UV}}(L)\sim L^{\alpha_g^{\text{UV}}}$ & $0.5\pm 0.1$ \\
		\hline $\alpha_l^{\text{IR}}$ & $C^{\text{IR}}(r)\sim r^{\alpha_l^{\text{IR}}}$ & $0$ \\
		\hline $\alpha_g^{\text{IR}}$ & $W^{\text{IR}}(L)\sim L^{\alpha_g^{\text{IR}}}$ & $0$ \\
		\hline $\gamma_{\sigma}$ & $\sigma_p \sim p^{-\gamma_{\sigma}}$ & $0.43\pm 0.03$ \\
		\hline $\gamma_{\beta}$ & $\beta \sim p^{-\gamma_{\beta}}$ & $3.19\pm 0.02$ \\
		\hline
	\end{tabular}
	\caption{The critical exponents of the local quantities.}
	\label{tab:local-exponents}
\end{table}

\subsection{geometrical properties}\label{global}

\begin{figure*}
	\centering
    \begin{subfigure}{0.49\textwidth}\includegraphics[width=\textwidth]{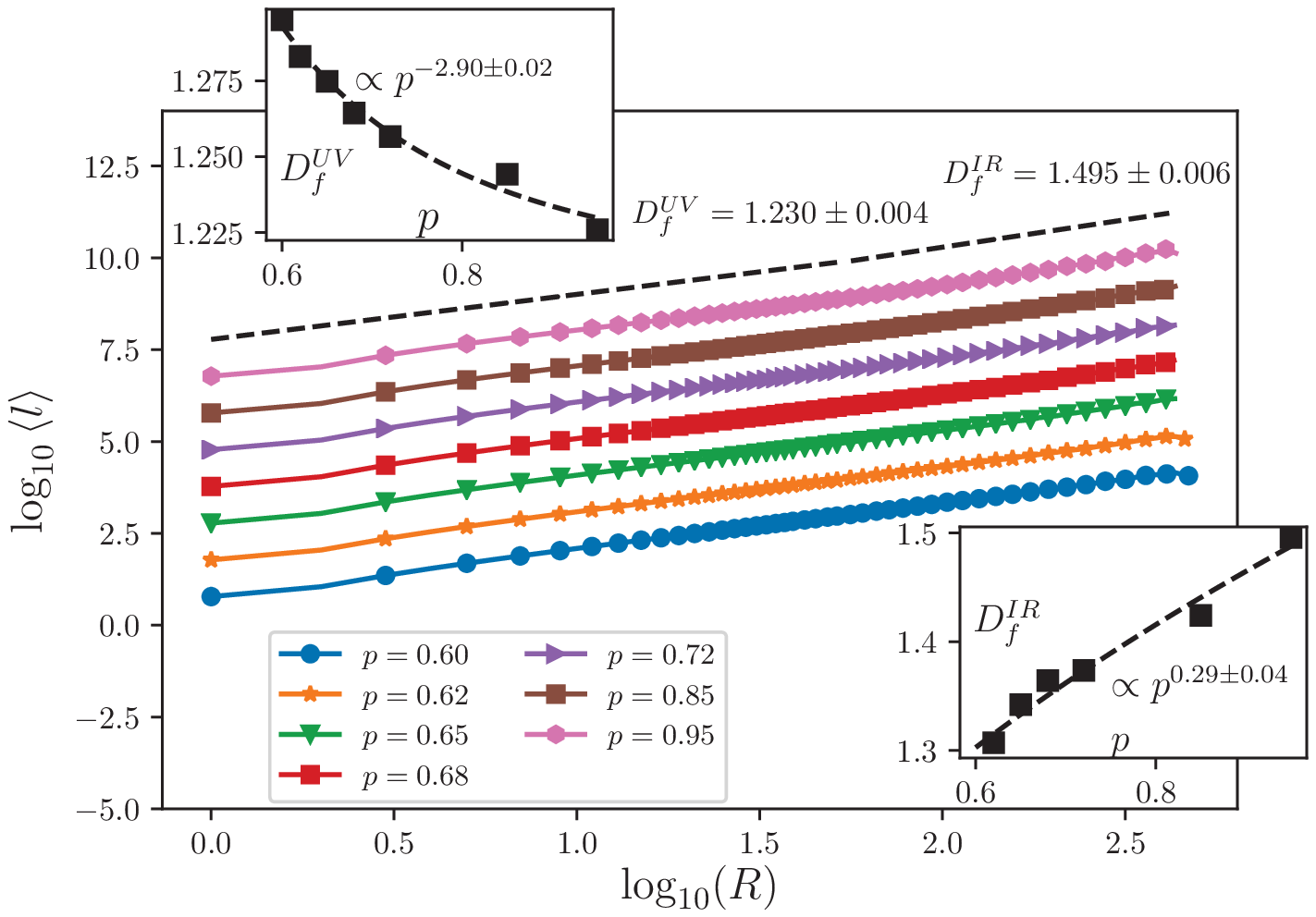}
		\caption{}
		\label{fig:Df}
	\end{subfigure}
	\begin{subfigure}{0.49\textwidth}\includegraphics[width=\textwidth]{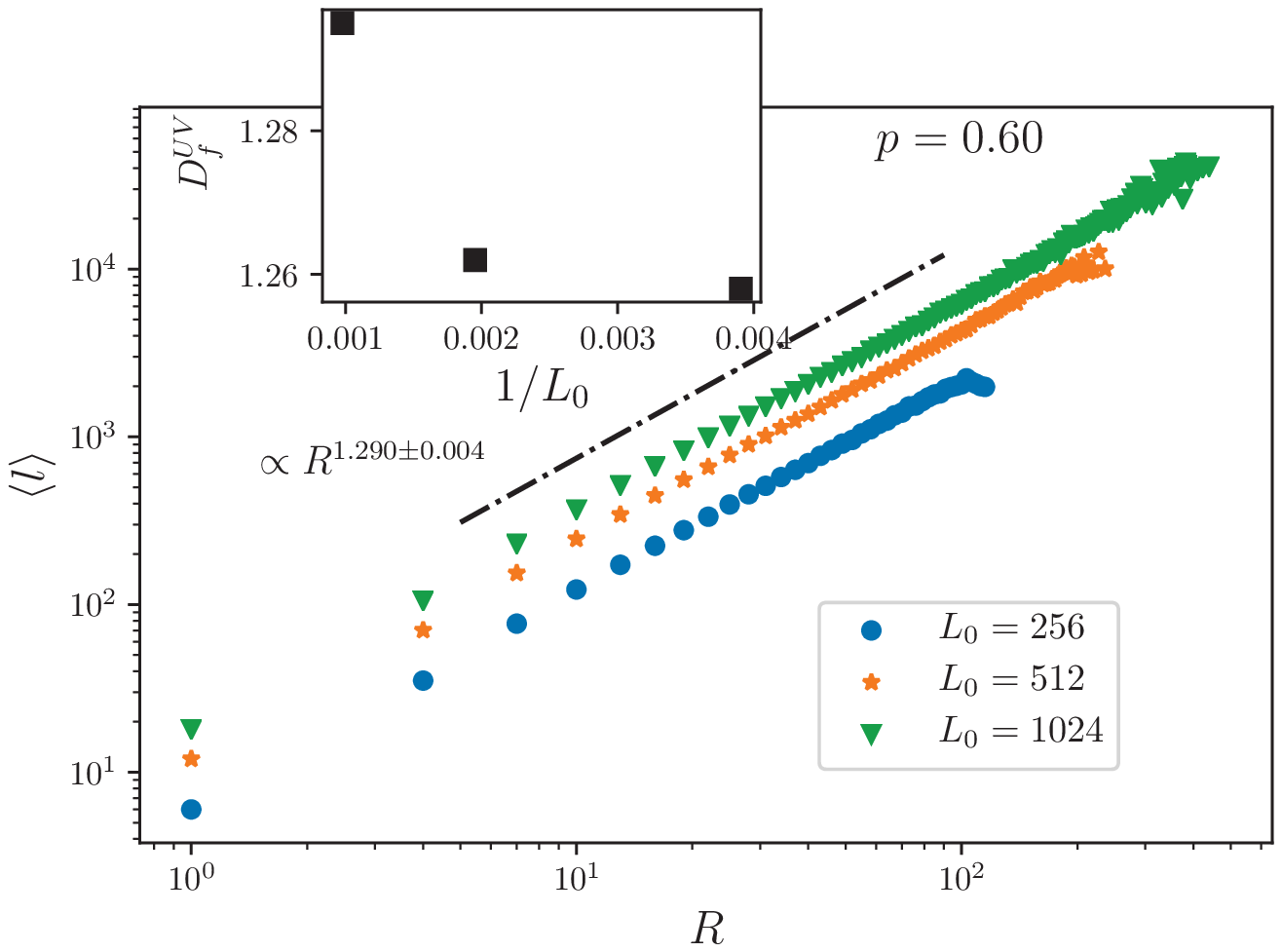}
		\caption{}
		\label{fig:L-Df}
	\end{subfigure}
	\caption{(Color online) (a) The fractal dimension defined by $\left\langle l\right\rangle\sim r^{D_f} $ for various rates of $p$ and $L_0=1024$. Upper inset: $D_f^{\text{UV}}$ in terms of $p$, Lower inset: $D_f^{\text{IR}}$ in terms of $p$. (b) The finite size dependence of the fractal dimension.}
	\label{fig:geometrical1}
\end{figure*}

The local features of the critical models imply some non-local properties, which make the problem worthy to be investigated from the geometrical point of view. This helps to distinguish more precisely the universality class of the model in hand. The example of such geometrical quantities can be found in Figs.~\ref{fig:pGreaterThanPcSolution} and~\ref{fig:pcSolution2} in which the level lines of the potential sample have been shown. By looking at these figures some observations can be made. For example we see that some large iso-potential islands have more chance to appear for $p=p_c$ when compared with $p>p_c$, e.g. $p=0.7$. \\
For the bulk regions, these level lines are some stochastic non-intersecting loops with different shapes and sizes. When a large number of potentials are obtained for various percolation samples, we have contour loop ensemble (CLE) for which many geometrical observables can be obtained. In this paper we analyze the length of loops ($l$), and the gyration radius of loops ($r$). For the mono-fractal systems one expects that the distribution function of these quantities show the power-law behavior, i.e. $P(x)\sim x^{-\tau_x}$ with $x=l$ and $r$ (which results from Eq.~\ref{eq:FSS}), and also $\left\langle l\right\rangle\sim r^{D_f} $ in which the exponent $D_f$ is the fractal dimension (FD) of loops. FD is a key geometrical exponent which is uniquely dedicated to universality classes of the critical systems and can be interpreted as the representative of that class. Also, according to the SLE theory, the 2D critical models are classified by means of the diffusivity parameter $\kappa$ (and the corresponding random curves are said to be SLE$_{\kappa}$) which is related to FD by the relation $D_f=1+\frac{\kappa}{8}$, which shows the importance of this exponent. For multi-fractal systems however these relations are not hold. For example, in many cases when there is a cross-over between two fixed points, two distinct critical exponents are observed~\cite{najafi2012avalanche}, for which Eq.~\ref{eq:FSS} is not applicable. In such cases, one may act just like the previous section, i.e. find the cross-over points and extract two exponents that is expected to be different for the two distinct regime.\\
Figure~\ref{fig:Df} contains the (shifted) $\log(\left\langle l\right\rangle )-\log(r)$ plot for $L_0=1024$ for various rates of $p$. Just like the local quantities in the previous section, here a smooth cross over takes place between two regimes: UV and IR, and the FD for these regimes are not the same. In two insets, we have shown $D_f^{\text{UV}}$ and $D_f^{\text{IR}}$ in terms of $p$ with some fitting line. A very similar fittings have been found for $L_0=256$ and $512$. By analyzing the upper inset, we conclude that $\lim_{p\rightarrow p_c} D_f^{\text{UV}}=1.295\pm 0.005$. This approach is of power-law form with exponent $2.90\pm 0.02$ which has been shown in the graph. In the lower inset however, we see that $\lim_{p\rightarrow 1} D_f^{\text{IR}}=1.50\pm 0.02$ and the corresponding exponent is $0.29\pm 0.04$. The obtained $D_f^{\text{IR}}(p\rightarrow 1)$ is just the FD of the level lines of Edwards-Wilkinson model, and also $\kappa=4$ SLE, i.e. $D_f=\frac{3}{2}$, which again confirms that the IR regime is described by GFF$_{p=1}$. In the Fig.~\ref{fig:L-Df} the log-log plot of $l-r$ has bee shown for $L_0=256,512$ and $1024$ for $p=0.6$. In the inset $D_F^{\text{UV}}$ has been shown in terms of the system size $L_0$ for the UV regime.\\

\begin{figure*}
	\centering
	\begin{subfigure}{0.49\textwidth}\includegraphics[width=\textwidth]{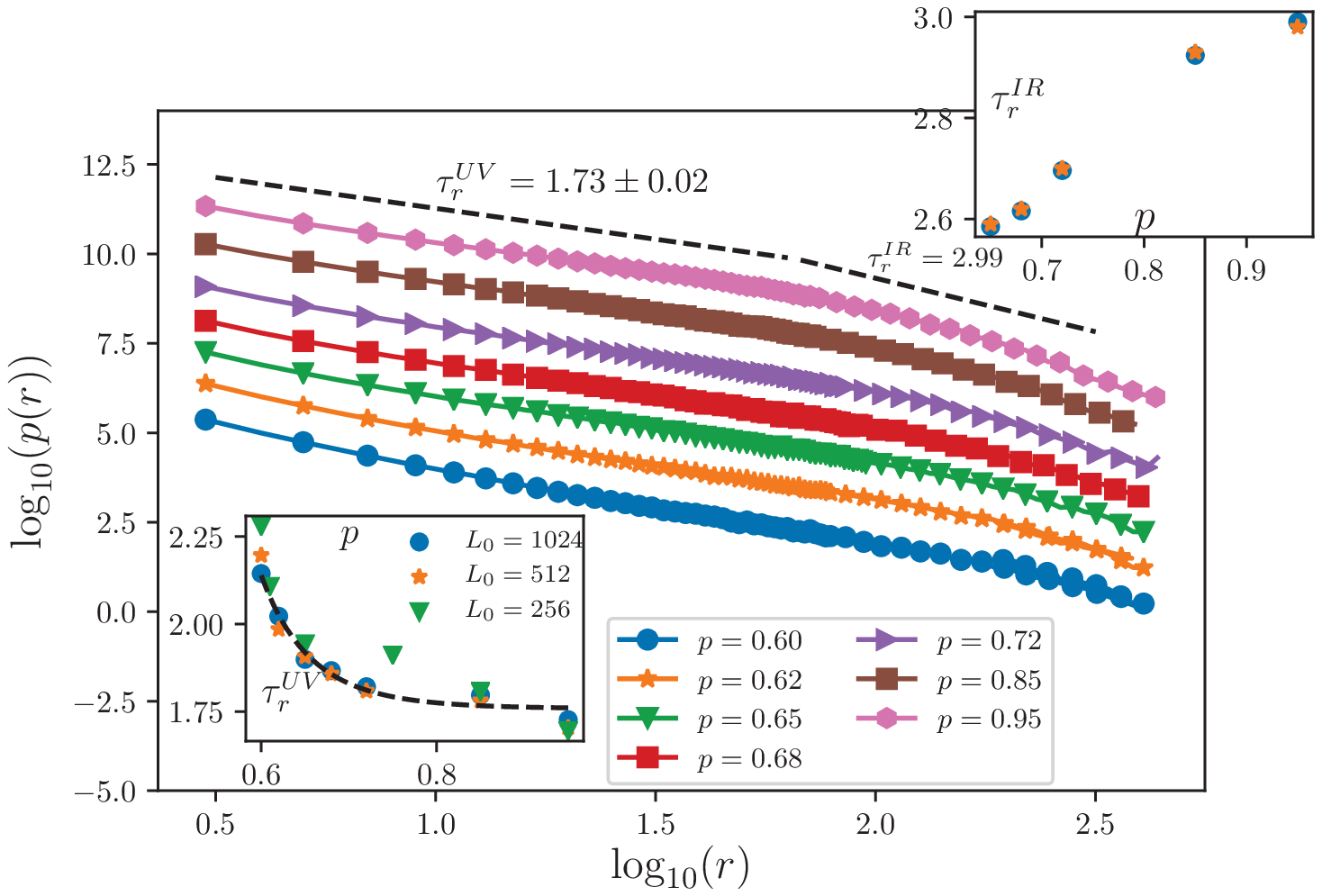}
		\caption{}
		\label{fig:p-r-L}
	\end{subfigure}
	\begin{subfigure}{0.49\textwidth}\includegraphics[width=\textwidth]{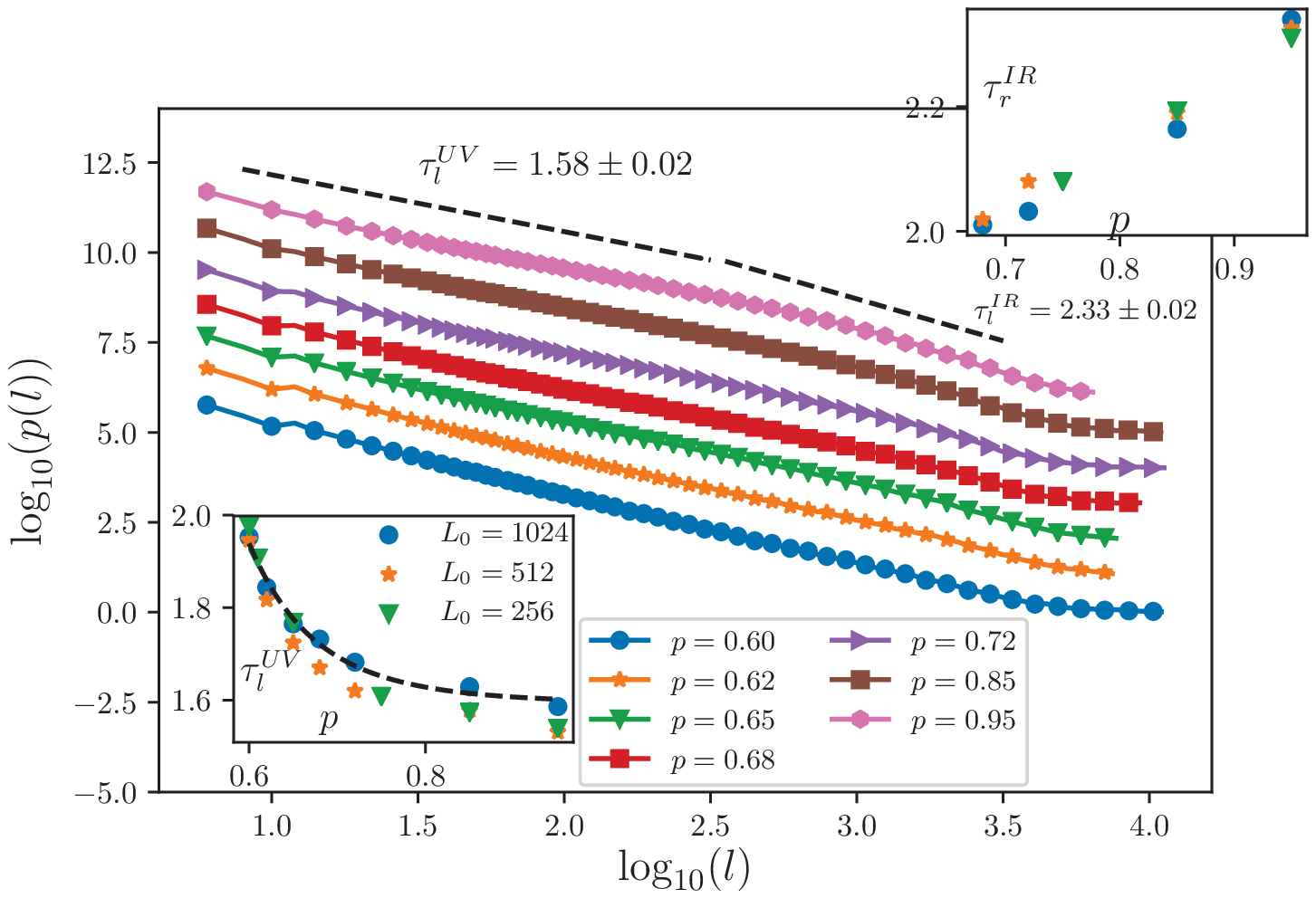}
		\caption{}
		\label{fig:p-l-L}
	\end{subfigure}
	\caption{(Color online) The log-log plot of the distribution function of (a) the gyration radius $r$, (b) the loop length $l$. The $p$ dependence of (c) $\tau_x^{\text{UV}}$ and $\tau_x^{\text{IR}}$, $x=r$ and $l$ have been shown in the insets. The trend-lines have been sketched for eyes helping.}
	\label{fig:geometrical2}
\end{figure*}

The other important data can be obtained by analyzing the distribution functions of $l$ and $r$, i.e. Figs~\ref{fig:p-r-L} and \ref{fig:p-l-L} with two-slope character. The upper insets justify the theoretical prediction of $\tau_r^{\text{GFF}}=3.0$ and $\tau_l^{\text{GFF}}=\frac{7}{3}$~\cite{kondev2000nonlinear}, i.e. $\lim_{p\rightarrow 1} \tau_r^{\text{IR}}=3.0\pm 0.1$ and $\lim_{p\rightarrow 1} \tau_l^{\text{IR}}=2.33\pm 0.02$. In the UV regime, some new exponents appear. From the lower insets we see that $\lim_{p\rightarrow p_c} \tau_r^{\text{UV}}=2.2\pm 0.1$ and $\lim_{p\rightarrow p_c} \tau_l^{\text{UV}}=1.95\pm 0.1$. These exponents have been gathered in TABLE~\ref{tab:global-exponents1}.\\

\begin{table}
	\begin{tabular}{c|c|c}
		\hline  & UV $p\rightarrow p_c$ regime & IR $p\rightarrow 1$ regime \\
	    \hline  $\tau_r$ & $2.2\pm 0.1$  &  $3.0\pm 0.1$  \\
	    \hline  $\tau_l$ & $1.95\pm 0.1$  &  $2.33\pm 0.03$  \\
	    \hline  $D_f$ & $1.295\pm 0.005$  &  $1.50\pm 0.02$  \\
		\hline
	\end{tabular}
	\caption{The critical exponents of the global quantities ($\tau_r$ and $\tau_l$) for $L=1024$.}
	\label{tab:global-exponents1}
\end{table}

\section*{Discussion and Conclusion}\label{conclusion}
\label{sec:conc}

\begin{figure*}
	\centerline{\includegraphics[scale=.40]{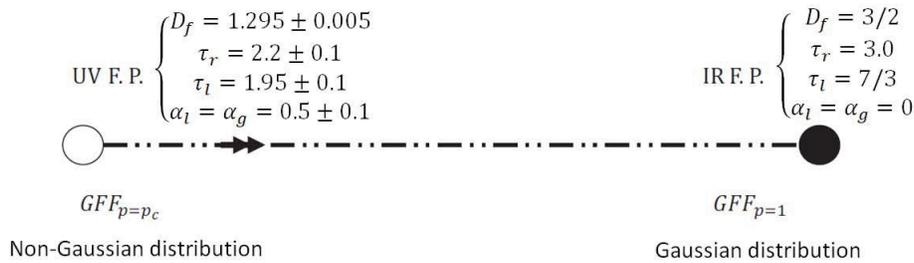}}
	\caption{(Color Online) The schematic representation of the structure of fixed points for the GFF$_p$ problem.}
	\label{fig:phasespace}
\end{figure*}

In this paper we have considered the Gaussian free field (GFF) in the 2D disordered media. To generate GFF samples in a regular lattice, one can consider the Edwards-Wilkinson (EW) model in the stationary state which is equivalent to the Poisson model in the background of white-noise charges. For disordering the host media, we have considered each site to be in one state of two possibilities: empty or occupied by metallic particle (namely the metallic site). The spatial arrangement of metallic particles was modeled by the percolation theory: with the probability $p$ the site is empty and with the probability $1-p$ the site is metallic. Therefore the media is composed of ordinary and metallic regions (inside which the potential is constant). We have mapped the problem to a rough random surface (with some iso-height islands) and have calculated the corresponding exponents. The most general finding of our (local) investigation has been that the incorporation of metallic particles to the system annihilates the spatial correlation of the potential field and also decreases the statistical fluctuations of them.\\
Two especial points were seen in the phase space: $p=p_c$ and $p=1$. The first point is called GFF$_{p=p_c}$ which captures the critical behaviors of the system in the small spatial scales (UV behaviors), and the last one is named as GFF$_{p=1}$ which captures the critical behaviors of the system in the large spatial scales (IR behaviors). We have detected a cross over region between these two limits around some spatial scale $r^*$ (corresponding to $l^*$), in which the behaviors smoothly changes from UV to IR region. By analyzing this point, we realized that, under enlarging the system size, the IR properties dominate the phase space, and equivalently the UV fixed point is unstable towards the IR fixed point which is described by the GFF in the regular system. On the other hand, the UV fixed point has new exponents which has been gathered in TABLEs~\ref{tab:local-exponents} and~\ref{tab:global-exponents1}. Accordingly we propose the phase space shown schematically in Fig.~\ref{fig:phasespace} which demonstrates the structure of fixed points. The hollow circle is representative of unstable (UV) fixed point with non-Gaussian distribution, and the solid circle shows the stable (IR) fixed point which is GFF in the regular system. The local and geometrical exponents of GFF$_p$ have been shown in this figure to facilitate comparison the fixed points in the future works.

We conclude that the dilution perturbation which is tuned by the off-criticality parameter $\epsilon_0\equiv 1-p$ is irrelevant for the $p=1$ fixed point. Also the unstable fixed point (GFF$_{p=p_c}$) which is composed of two ingredients ($c=1$ and $c=0$ CFTs, the former as the dynamical model and the latter as the host media) contains some critical exponents (importantly $D_f=1.295\pm 0.005$, corresponding to $\kappa=2.36\pm 0.04$) and should be characterized in some more details.

\bibliography{refs}

\end{document}